\documentclass[conference]{IEEEtran}
\usepackage{graphicx}
\usepackage{subfigure}
\usepackage{float}
\usepackage{amsmath}
\usepackage{amsfonts,amssymb}
\usepackage{dsfont}
\usepackage{algpseudocode}
\usepackage{makecell}
\usepackage{cite}
\usepackage{multirow}
\usepackage{multicol}
\usepackage{booktabs}
\usepackage[ruled,vlined, linesnumbered,lined,boxed,commentsnumbered,ruled,longend,noend]{algorithm2e}
\usepackage{xcolor,soul}
\usepackage{comment}
\usepackage{bbm}

\SetKwInput{KwInput}{Input}                
\SetKwInput{KwOutput}{Output}              

\usepackage{fancyhdr,lipsum}

\fancypagestyle{mahmood}{%
   \fancyhf{} 
   
   \fancyhead[C]{You can use this material personally. Reprinting or republishing this material for the purpose of advertising or promotion, creating new collective works, reselling or redistributing to servers or lists, or using any copyrighted component in other works must adhere to IEEE policy. The DOI will be supplied as soon as it becomes available.}
}%

\makeatletter
\let\ps@IEEEtitlepagestyle\ps@mahmood
\makeatother

\begin{document}
\title{A Semantic-Aware Multiple Access Scheme for Distributed, Dynamic 6G-Based Applications}

\author{
    \IEEEauthorblockN{
        Hamidreza Mazandarani\textsuperscript{1}, Masoud Shokrnezhad\textsuperscript{2}, and Tarik Taleb\textsuperscript{1, 2} \\
    }
    \IEEEauthorblockA{
       \textsuperscript{1} \textit{Ruhr University Bochum, Bochum, Germany; hr.mazandarani@ieee.org, tarik.taleb@rub.de} \\
        \textsuperscript{2} \textit{Oulu University, Oulu, Finland; \{masoud.shokrnezhad, tarik.taleb\}@oulu.fi}
    }
}

\maketitle

\begin{abstract}
The emergence of the semantic-aware paradigm presents opportunities for innovative services, especially in the context of 6G-based applications. Although significant progress has been made in semantic extraction techniques, the incorporation of semantic information into resource allocation decision-making is still in its early stages, lacking consideration of the requirements and characteristics of future systems. In response, this paper introduces a novel formulation for the problem of multiple access to the wireless spectrum. It aims to optimize the utilization-fairness trade-off, using the $\alpha$-fairness metric, while accounting for user data correlation by introducing the concepts of self- and assisted throughputs. Initially, the problem is analyzed to identify its optimal solution. Subsequently, a Semantic-Aware Multi-Agent Double and Dueling Deep Q-Learning (SAMA-D3QL) technique is proposed. This method is grounded in Model-free Multi-Agent Deep Reinforcement Learning (MADRL), enabling the user equipment to autonomously make decisions regarding wireless spectrum access based solely on their local individual observations. The efficiency of the proposed technique is evaluated through two scenarios: single-channel and multi-channel. The findings illustrate that, across a spectrum of $\alpha$ values, association matrices, and channels, SAMA-D3QL consistently outperforms alternative approaches. This establishes it as a promising candidate for facilitating the realization of future federated, dynamically evolving applications.
\end{abstract}


\section{Introduction}
\begin{figure}[t!]\centering
\includegraphics[width=3.3in, height=1.5in]{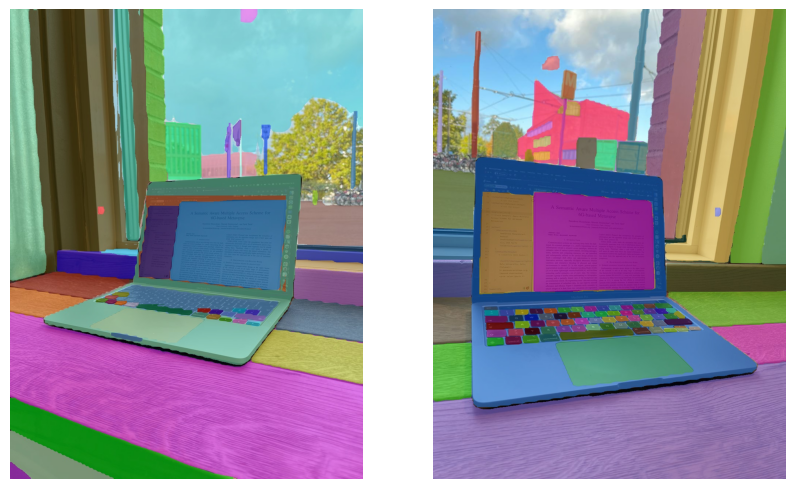}
    \vspace{-5pt}
    \caption{An illustration of similar segments.}
    \vspace{-20pt}
    \label{sample_sam}
\end{figure}

Communication systems are undergoing a transformation from the traditional \textit{bit-oriented} model, focused solely on transmitting data bits, to a \textit{semantic-aware} paradigm. In this new approach, the semantic value of various system components (such as messages, users, and resources) is quantified and subsequently integrated into the service provisioning process \cite{our_mag_paper}. This paradigm shift opens up significant opportunities for innovative services, particularly in the context of the 6G-based Metaverse, where seamless integration of physical and virtual environments is a key goal \cite{haoyu2023xr}. To support this transition, versatile data segmentation methods like the Segment Anything Model (SAM) \cite{kirillov2023segment} and its lightweight version FastSAM \cite{zhao2023fast} play a pivotal role in capturing meaningful understanding from user data. For instance, consider a scenario involving a group of nearby users engaged in a holographic meeting with a partially shared background and environmental sounds. Through data segmentation and semantic extraction, it becomes feasible to improve network efficiency and effectiveness and accommodate a greater number of users and requests with stringent end-to-end requirements. Fig. \ref{sample_sam} demonstrates this concept, showcasing segmented images of a desk from different viewpoints, revealing shared similarities across multiple segments.

The incorporation of semantics into decision-making processes at various network layers is still in its nascent stages of development. Our knowledge management framework for semantic-aware systems, known as KB MANO \cite{our_mag_paper}, in conjunction with the introduction of the knowledge space concept within the Meta-Networking framework \cite{lin2023meta} and the reasoning plane discussed in \cite{chaccour2022less}, represents some of the initial endeavors to conceptualize network-wide semantic awareness. Particularly, in the data link layer, a semantic-aware Medium Access Control (MAC) protocol can optimize the utilization of limited radio resources by eliminating redundant data transmission. The Model Division Multiple Access (MDMA) is an example of such an approach \cite{zhang2023model}, dedicating a portion of resources to transmit shared information among users. In a similar vein, Kassab \textit{et al.} \cite{kassab2020multi} conducted research focusing on a scenario involving a collective of Internet of Things (IoT) devices engaged in the concurrent monitoring of correlated on-off events, such as variations in temperature levels. Another line of research, exemplified by DeepSC-IR/MT \cite{xie2022task} and DeepMA \cite{zhang2023deepma}, explores the joint semantic encoding of data from multiple users (i.e., multi-user semantic communications). While these methods are effective in specific contexts, they do not account for the distributed and dynamic nature of future applications, making them less scalable and potentially unsuitable for 6G systems.

When dealing with an environment containing ever-changing multiple distributed actors, Multi-Agent Deep Reinforcement Learning (MADRL) stands out as an effective approach. Its adaptability to handle non-stationarity, stemming from the co-evolution of agents' policies, makes it a compelling choice \cite{gronauer2022multi}. In MADRL, agents base their transmission decisions on historical observations, such as success or failure in task fulfillment, while coordination among agents occurs through policy sharing, message passing, or centralized training. MADRL-based multiple access control has been discussed in various papers. For instance, Naparstek \textit{et al.} \cite{naparstek2018deep} proposed an efficient Nash equilibrium for multi-agent MAC where all agents adopt the same policy. Sohaib \textit{et al.} \cite{sohaib2021dynamic} extended MADRL to MAC by using transmission policies over consecutive time slots as decision variables (as opposed to per-time-slot decision-making), ensuring short-term fairness under varying user conditions. Guo \textit{et al.} \cite{guo2022multi} introduced the QMIX-advanced LBT (QLBT) MAC protocol, building upon the QMIX algorithm for MADRL problems \cite{rashid2020monotonic}. QLBT incorporates two reward signals, one for maximizing network utilization and another for maintaining fairness. However, despite these advancements, none of the examined MADRL-based MAC protocols explicitly address the growing need for semantic awareness, a critical aspect of evolving communication systems.

In response to the existing gap in the current literature, we propose a novel formulation for the multiple-access problem to optimize the trade-off between utilization and fairness, taking into account the inherent correlation in users' data. This departure from the conventional bit-oriented paradigm underscores our paradigm shift towards semantic awareness. Notably, we demonstrate the intractable nature of this problem and present a solution approach employing model-free MADRL. Our contribution, the Semantic-Aware Multi-Agent Double and Dueling Deep Q-Learning (SAMA-D3QL) method, expanding upon the principles established in our prior research \cite{our_cl_d3ql_paper, mazandarani2023self, farhoudi_qos-aware_2023, shokrnezhad_double_2023}. SAMA-D3QL operates under the premise that users possess dynamic associations with various semantic segments, some of which may be shared among them. Consequently, when a user transmits a packet, it benefits all users with whom she shares pertinent segments.


The subsequent sections of this paper are organized as follows. Section \ref{s_pf} introduces the system model and provides a comprehensive problem formulation. The optimal problem solution and the details of the SAMA-D3QL technique are elucidated in Section \ref{s_pm}. Section \ref{s_sim} presents the numerical results, while Section \ref{s_con} concludes our work with closing remarks.

\section{Problem Formulation}\label{s_pf}
In this study, a configuration comprising a Small Base Station (SBS) and a collection of $\mathcal{N}$ intelligent User Equipment (UE) is examined, each uniquely labeled as $u_{i}$, where $i \in \mathbb{N} = \{1, \ldots, \mathcal{N}\}$. These UEs are engaged in contention for access to $\mathcal{C}$ perfectly time-slotted communication channels designated for transmitting their data to the SBS. It is essential to note that simultaneous transmissions over a single channel inevitably lead to collision. Consequently, in the conventional bit-oriented framework, the highest achievable network utilization is capped at $\mathcal{C}$ packets per time slot.

\begin{figure}[t!]\centering
\includegraphics[width=2.7in]{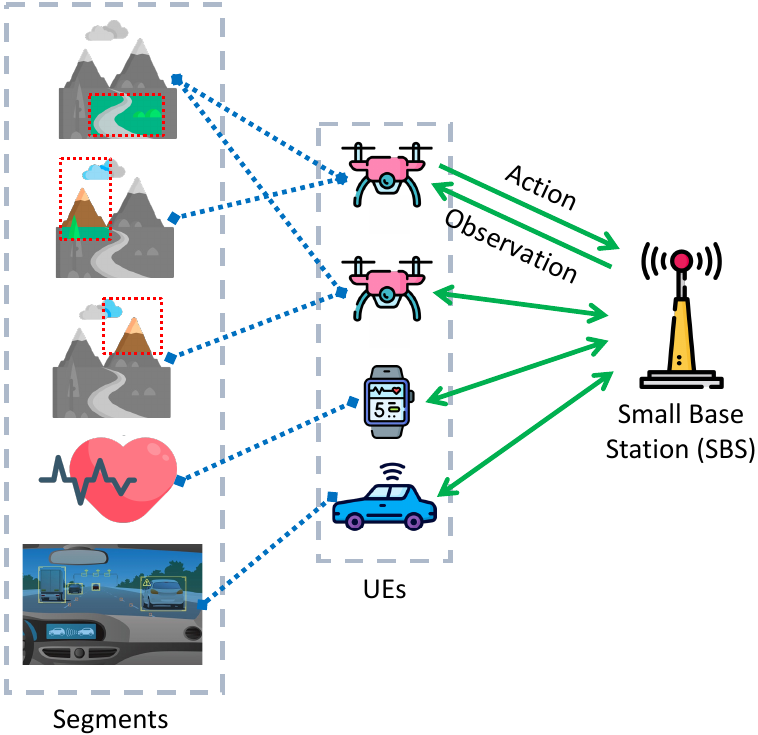}
    \vspace{-5pt}
    \caption{A sample network consisting of four UEs ($\mathcal{N} = 4$) and five segments ($\mathcal{K} = 5$). As illustrated, one separate segment is associated with each user, in addition to one shared segment among the first and second UEs.}
    \vspace{-15pt}
    \label{fig_sample_net}
\end{figure}

It is presumed that within the network, there exists shared semantic information among subsets of UEs. For instance, in a holographic meeting, nearby UEs may share some environmental data while retaining their individual face and voice data. This scenario is represented by $\mathcal{K}$ distinct semantic entities, termed \textit{segments}, denoted as $k \in \mathbb{K} = \{1, \ldots, \mathcal{K} \}$. At each time slot $t \in \{0, \ldots, \mathcal{T} \}$, each UE $i$ is associated/disassociated with a segment $k$, signified by a binary variable $a_{i,k}^{t} \in \{0, 1\}$, forming a predefined binary association matrix $\mathbb{A} = [a_{i,k}^{t}]_{\mathcal{N} \times \mathcal{K} \times \mathcal{T}}$. The matrix serves as the input to the system, and an orchestration entity, such as KB MANO as discussed by Shokrnezhad \textit{et al.} \cite{our_mag_paper}, is responsible for generating it. In an illustrative example, shown in Fig. \ref{fig_sample_net}, the association matrix at time slot $t$ is given as $\mathbb{A}^{t} = \big[[1, 1, 0, 0, 0],[1, 0, 1, 0, 0],[0, 0, 0, 1, 0],[0, 0, 0, 0, 1]\big]$.

To evaluate the semantic efficiency, we introduce a segment transmission indicator, denoted as $y_{k}^{t}$. This indicator operates under the assumption that, at any given time slot, all UE-associated segments are combined into a single packet. Furthermore, receiving a specific segment by the SBS multiple times within a time slot does not enhance throughput. $y_{k}^{t}$ is defined as:
\begin{equation}\label{segment_indicator}
y_{k}^{t} \triangleq \min \Bigg\{ 1, \sum_{ j \in \mathbb{N} }^{}{a_{j,k}^{t} \cdot z_{j}^{t}} \Bigg\},
\end{equation}
where $z_{j}^{t} \in \{0,1\}$ signifies whether UE $j$ has successfully transmitted a packet during time slot $t$. In Appendix \ref{app_a}, a more generalized form is presented that does not rely on the assumption that receiving multiple copies of a segment by the SBS is unproductive.

Based on \eqref{segment_indicator}, the normalized throughput of each UE can be determined. Considering that $\mathbb{W}^{t} = \{ t-\mathcal{H}, \ldots, t \}$ represents a sliding time window of length $\mathcal{H}$, and ${\Omega}_{i}^{t} = 1/\big(\sum_{k, t \in \mathbb{K}, \mathbb{W}^{t}} {a_{i,k}^{t}}\big)$ acts as the normalization factor, the normalized throughput of $u_{i}$ is as follows:
\begin{equation}\label{ue_throughput}
x_{i}^{t} \triangleq {\Omega}_{i}^{t} \cdot \sum\limits_{k, t \in \mathbb{K}, \mathbb{W}^{t}} {a_{i,k}^{t} \cdot y_{k}^{t}} \\
\end{equation}
By isolating $u_{i}$ from other UEs in \eqref{ue_throughput}, we can break down $x_{i}^{t}$ into two components: \textit{self throughput} ($\dot{x}_{i}^{t}$) and \textit{assisted throughput} ($\ddot{x}_{i}^{t}$), defined in \eqref{ue_throughput_decomposed}. The assisted throughput for UE $i$ represents the contribution of other UEs in transmitting shared segments between them and $u_i$ to the SBS.
\begin{align}\label{ue_throughput_decomposed}
    & {x}_{i}^{t} = \dot{x}_{i}^{t} + \ddot{x}_{i}^{t} \notag \\
    & \dot{x}_{i}^{t} = {\Omega}_{i}^{t} \cdot \sum\limits_{k, t \in \mathbb{K}, \mathbb{W}^{t}} {a_{i,k}^{t} \cdot  z_{i}^{t}} \\
    & \ddot{x}_{i}^{t} = \ {\Omega}_{i}^{t} \cdot \sum\limits_{k,t \in \mathbb{K}, \mathbb{W}^{t}} { \Big( a_{i,k}^{t} \cdot \ (1 - z_{i}^{t}) \cdot y_{k}^{t, -i} \Big) } \notag
\end{align}
Here, $y_{k}^{t, -i} = \min \big\{ 1, \sum_{ j \in \mathbb{N} \text{\textbackslash} \{ i \} }^{}{a_{j,k}^{t} \cdot z_{j}^{t}} \big\}$.

\vspace{5px}

The problem can now be formally defined. Our objective is to maximize network utilization across the network's lifetime while maintaining fairness among UEs. We adopt the $\alpha$-fairness metric as the objective function, as expressed in C1. With $\alpha = 0$ and $\alpha = \infty$ as two extreme cases: sum throughput maximization and max-min fairness maximization, respectively. In our time-slotted multi-channel network configuration, the successful transmission of $u_{i}$ at time slot $t$ (i.e., $z_{i}^{t}$) is contingent upon $u_{i}$ transmitting a packet on a channel that all other UEs are not utilizing. C2 and C3 formalize this condition, where ${m}_{i, c}^{t}$ denotes whether $u_{i}$ is transmitting on channel $c$ at time slot $t$. C2 ensures that each UE transmits on at most one channel during each time slot, and C3 guarantees that $z_i^t$ is equal to $1$ if $u_i$ selects a channel that is not already chosen by other users. Consequently, the transmission matrix $\mathbb{M} = [{m}_{i, c}^{t}]_{\mathcal{N} \times \mathcal{C} \times \mathcal{T}}$ determines $\mathbb{Z} = [z_i^t]_{\mathcal{N}\times \mathcal{T}}$, which influences UE throughputs $\mathbb{X} = [x_i^t]_{\mathcal{N}\times \mathcal{T}}$ as per \eqref{ue_throughput_decomposed}. Finally, we calculate $\mathbf{U}_{\alpha}(\mathbb{X})$. In symbolic terms, we have a causal relationship denoted as $\mathbb{M} \rightarrow \mathbb{Z} \rightarrow \mathbb{X} \rightarrow \mathbf{U}_{\alpha}(\mathbb{X})$, where the arrow indicates the causal flow.
\begin{align}\label{problem}
    & \max_{\mathbb{M}} \sum_{t \in \{0, \ldots, \mathcal{T} \}} \mathbf{U}_{\alpha}(\mathbb{X}^{t}) \quad \mbox{s.t. \ } \\
    & \mathbf{U}_{\alpha}(\mathbb{X}^{t}) = 
    \begin{cases} 
        \sum\limits_{i \in \mathbb{N}}{ log(x_{i}^{t})} & \text{if } \alpha=1 \\
        (1 - \alpha)^{-1} \cdot \sum\limits_{i \in \mathbb{N}}{ {(x_{i}^{t})}^{1-\alpha}} & \text{if } \alpha \neq 1
    \end{cases} \tag{C1} \\
    & \sum_{c \in \{0, \ldots, \mathcal{C} \}}{{m}_{i, c}^{t}} \leq 1 \tag{C2} \\
    & z_{i}^{t} = \sum_{c \in \{0, \ldots, \mathcal{C} \}}{{m}_{i, c}^{t} \cdot \prod_{j\neq i}^{}{(1 - {m}_{j, c}^{t}}} ) \tag{C3} \\
    & x_{i}^{t} = \dot{x_{i}}^{t} + \ddot{x_{i}}^{t} \quad \text{according to \eqref{ue_throughput_decomposed}} \tag{C4}
\end{align}
Note that C2, C3, and C4 are applicable to all values of $i \in \mathbb{N}$ and $t \in \{0, \ldots, \mathcal{T} \}$.

\section{Proposed Method}\label{s_pm}
\subsection{Optimal Solutions}
In a semantic-oblivious system (where each UE has its segments and no assisted throughputs exist), it is proved that the optimal point for \eqref{problem} is attained when all UEs receive the same channel allocation ($x_i^t = \min\{\mathcal{C} / \mathcal{N}, 1\}$ for all $i$ and $t$) \cite{naparstek2018deep}. However, in a semantic-aware system, due to the inter-dependency of UEs' throughputs, achieving optimality is not as straightforward as allocating resources equally. Returning to the modest example of Fig. \ref{fig_sample_net}, optimal allocations for different $\alpha$ values are provided in Table \ref{sample_net_optimal} based on an exhaustive search over the solution space (assuming $\mathcal{C} = 1$ and stationarity over time). The first row, representing no fairness ($\alpha = 0$), allocates the entire channel to the first two UEs who share each other's information, resulting in an optimal objective value of $1.5$, a 50\% improvement over the bit-oriented scenario. As $\alpha$ increases, the last two UEs gain a larger share of the channel as there is no assisted throughput for them.

In more intricate and realistic settings, characterized by an increased number of UEs, channels, or segments, the optimization of the nonlinear problem presented in \eqref{problem} becomes challenging, if not infeasible. Furthermore, it is essential to acknowledge that, in real-world scenarios, we do not have access to the information of future time slots. Additionally, in distributed scenarios where UEs are unaware of each other's decisions or associations, we face a fundamental information gap that hinders optimal problem-solving. Consequently, we turn to model-free MADRL approaches, which will be elaborated upon in the following subsection.

\begin{table}
\centering
\caption{The optimal allocations for the instance depicted in Fig. \ref{fig_sample_net}.}
\vspace{-5pt}
\begin{tabular}{|c|c|c|}
\hline
\textbf{$\alpha$} & \textbf{Self Throughputs} & \textbf{Normalized Throughputs} \\
\hline
0 & $ [0.5, 0.5, 0, 0] $ & $ [0.75, 0.75, 0, 0] $ \\
0.5 & $ [0.3, 0.3, 0.2, 0.2] $ & $ [0.45, 0.45, 0.2, 0.2] $ \\
1 & $ [0.25, 0.25, 0.25, 0.25] $ & $ [0.375, 0.375, 0.25, 0.25] $ \\
$\infty$ & $ [0.2, 0.2, 0.3, 0.3] $ & $ [0.3, 0.3, 0.3, 0.3] $ \\
\hline
\end{tabular}
\vspace{-10pt}
\label{sample_net_optimal}
\end{table}

\subsection{Learning Mechanism} \label{learn_mech}
In what follows, we will elucidate the fundamental components of our MADRL-based approach, denoted as Semantic-Aware Multi-Agent Double and Dueling Deep Q-Learning (SAMA-D3QL). This exposition will encompass discussions on action and state spaces, reward mechanisms, and the architectural framework of the learning process.

\subsubsection{Action Space}
The action space for all UEs is specified in \eqref{action_space}. In this set, the action of $u_{i}$ can take one of two forms: ${\rho}_{i}: (0, c)$, representing the act of sensing channel $c$, or ${\rho}_{i}: (1, c)$, signifying the transmission of a packet on channel $c$.
\begin{equation}\label{action_space}
\boldsymbol{\rho} = { \bigg\{ \big\{ {\rho}_{i}: (\phi, c) | \phi \in \{0, 1\}, c \in \{1, ..., \mathcal{C}\} \big\} \ \bigg| \ i \in \mathbb{N} \bigg\} }
\end{equation}

\subsubsection{State Space}
The state space consists of four distinct components. To provide UEs with real-time fairness information, we incorporate the \textit{delay to last successful transmission} (D2LT) metric \cite{guo2022multi} represented as $v_{i}$. This metric signifies the number of time slots elapsed since the last successful transmission by $u_{i}$. Specifically, we include the normalized D2LT (i.e., $ \bar{v}_{i} = v_{i}/ \sum_{j \in \mathbb{N}}{v_{j}}$) in the state space of UEs. The following component relates to the UE's observation. When sensing channel $c$, the observation set for each UE is denoted as $\dot{o}_{i}$ = \{\textit{B: Busy, I: Idle}\}, while in the case of packet transmission, it is $\dot{o}_{i}$ = \{\textit{S: Success, C: Collision}\}. Furthermore, we introduce another indicator called "assisted transmission," denoted as $\ddot{o}_{i} \in [0, 1]$. At each time slot, the assisted transmission of $u_{i}$ is simply the ratio of its segments transmitted by other UEs. Given that the final element within the state space pertains to UE actions, the state of each UE comprises the most recent $\mathcal{H}$ (D2LT, observation, assisted transmission, action) tuples. Thus, the global system state at time slot $t$ is defined as follows:
\begin{equation}\label{state_space}
\boldsymbol{S}^{t} = { \bigg\{ {s}_{i}^{t} : \Big\{(\bar{v}_{i}^{t}, \dot{o}_{i}^{t}, {\ddot{o}_{i}}^{t}, {\rho}_{i}^{t}) | t \in \mathbb{W}^{t} \Big\} \ \bigg| \ i \in \mathbb{N} \bigg\} }
\end{equation}

\subsubsection{Reward}
Since we consider a cooperative scenario in which all UEs aim to maximize $\alpha$-fairness of the network, the system reward is empirically defined as the weighted average of successful transmissions, as outlined in \eqref{reward}. These weights are determined by the D2LT values, encouraging UEs that have experienced longer delays to access the channel.
\begin{equation}\label{reward}
\boldsymbol{R}^{t} = \frac{1}{\mathcal{N}} \ \cdot \sum_{i \in \mathbb{N}} {\bar{v}_{i}^{t} \ \cdot \mathbbm{1}(\dot{o}_{i}^{t} = S)} 
\end{equation}

\subsubsection{Training Process}
Our proposed methodology is founded on the Value Decomposition Network (VDN) framework \cite{sunehag2017value}, which adheres to the Centralized Training and Decentralized Execution (CTDE) paradigm. This strategy operates on the premise that during the training phase, full access to the global system state is available, while user decisions are made based on their local information. Similarly, the training process of our proposed method is centralized within the SBS, where UE policies are being trained. Given that the SBS possesses knowledge of UE experiences, including their actions and observations, there is no need for agents to transmit their experiences to the SBS. However, it is assumed that the SBS periodically updates and disseminates policies to the UEs via high-bandwidth and low-latency dedicated communication channels. UEs autonomously select their actions based on their respective policies and individual observations without relying on the actions or states of other UEs. Utilizing Q-learning, the summation of individual user Q-values is employed to compute the total Q-value within VDN, precisely calculated using the following equation:
\begin{align}\label{training}
Q_{tot}(\boldsymbol{S}^{t}, \boldsymbol{\rho}^{t}) = \sum_{i \in \mathbb{N}}{}{Q_{i}({s}_{i}^{t}, \rho_{i}^{t})}
\end{align}
With the above definition, the policy parameter update function for all agents, represented as $\boldsymbol{\mathcal{W}}  = \{ \mathcal{W}_{i} \ | \ i \in \mathbb{N} \}$, can be expressed as follows:
\begin{align}
\label{eq_DQL_bellman}
\boldsymbol{\mathcal{W}}^{t+1} = & \boldsymbol{\mathcal{W}}^{t} \nonumber \ + \\
& \sigma[Y^t_{D3QL} - Q_{tot}(\boldsymbol{S}_{t}, \boldsymbol{\mathbb{\rho}}_{t}; \boldsymbol{\mathcal{W}}^t)]\nabla_{\boldsymbol{\mathcal{W}}^{t}} \nonumber \ \cdot \\
& Q_{tot}(\boldsymbol{S}_{t}, \boldsymbol{\mathbb{\rho}}_{t}; \boldsymbol{\mathcal{W}}^{t}),
\end{align}
where the target value ($Y^t_{D3QL}$) is determined using the D3QL algorithm, which is an extension of the original DQL algorithm that incorporates both dueling and double mechanisms \cite{our_cl_d3ql_paper}. Our approach is outlined in Algorithm \ref{alg_sama_d3QL}, wherein lines 4-15 pertain to the decentralized execution of actions by UEs, while lines 17-20 depict the centralized training conducted at the SBS.

\begin{algorithm}[t!]\label{alg_sama_d3QL}
\caption{SAMA-D3QL}
\KwInput{$\mathcal{T}$, $\epsilon'$, and $\widetilde{\epsilon}$}
$\boldsymbol{\mathcal{W}} \leftarrow \mathbf{0}$, $\boldsymbol{\mathcal{W}}^{-} \leftarrow \mathbf{0}$, $\epsilon \gets 1$, $memory \gets \{\} $\\
\ForEach{$t$ in $\{0, \ldots, \mathcal{T} \}$}
{
    \textcolor{gray}{$\star$ Decentralized Execution (4-15)}  \\
    \ForEach{$i$ in $\mathbb{N}$}
    {
        $\zeta \gets$ generate a random number from $[0:1]$ \\
        \If{$\zeta > \epsilon$}
        {
            $(\phi, c) \gets$ argmax$_{\rho \in \boldsymbol{\rho}_{i}} Q({s}_{i}^{t}, \rho, {\mathcal{W}}_{i})$ \\
        }
        \Else
        {
            select a random $(\phi, c)$ from $\boldsymbol{\rho}_{i}$
        }
        \If{$\phi = 1$}
        {
            transmit packet on channel $c$
        }
        \Else
        {
            sense channel $c$
        }
        receive $\dot{o}_{i}^t$ and $\ddot{o}_{i}^t$ from the SBS \\
        construct ${s}_{i}^{t+1}$ \\
    }
    \textcolor{gray}{$\star$ Centralized Training (17-20)} \\
    calculate $\boldsymbol{R}^{t}$ according to \eqref{reward} \\
    $memory \gets \{ \boldsymbol{R}^{t} \} \cup \{({s}_{i}^{t}, (\phi, c), {s}_{i}^{t+1}) \big| \ i \in \mathbb{N} \}$ \\
    choose a batch of samples from $memory$\\
    train the agent according to \eqref{eq_DQL_bellman} \\
    \If{$\epsilon > \widetilde{\epsilon}$}
    {
        $\epsilon \gets \epsilon \cdot \epsilon'$
    }
}
\end{algorithm}

\section{Evaluation}\label{s_sim}
In this section, a series of experiments is conducted to perform a numerical evaluation of SAMA-D3QL. Each experiment involves a comparison between SAMA-D3QL and its semantic-oblivious counterpart (in terms of both network and algorithmic design), referred to as MA-D3QL. Specifically, MA-D3QL does not incorporate assisted transmission into its decision-making process and operates under the assumption that the SBS lacks awareness of the correlation among UEs' source data, thereby decoding only the self-throughputs. Additionally, we compare the results with those obtained using random agents in the semantic-aware network, serving as a benchmark. To facilitate algorithm comparison, we vary the alpha value in the $\alpha$-fairness function, considering four settings: $0$, $0.5$, $1$, and $2$, representing different degrees of fairness. In the first scenario, we establish two modest association matrices, for which optimal values can be extracted, to delve into the effectiveness of our strategy in detail. Subsequently, in the second scenario, we assess the performance of our scheme in a multi-channel setting involving a higher number of nodes and segments. Table \ref{tab_sim_par} details the simulation's hyperparameters and configurations. It is noteworthy that we have opted for LSTM networks as the feature extractor, owing to their capability to process sequential data.

\begin{table}[t!]
\caption{Training Configurations.}
\begin{center}
\vspace{-15pt}
\begin{tabular}{|c|c|}
\hline
\textbf{Parameter} & \textbf{Value} \\
\hline
State size ($\mathcal{H}$) & $4$ experiences \\
Capacity of experience memory & $500$ experiences \\
Batch size & $32$ \\
Discount Factor ($\gamma$) & 0.9 \\
Learning rate & $0.001$ \\
Exploration parameters $\widetilde{\epsilon}$, $\epsilon'$ & 0.005, 0.995 \\
Approximator model & \begin{tabular}{@{}c@{}}LSTM with $64$ units + \\ fully connected with $64$ and $32$ units \end{tabular}  \\
\begin{tabular}{@{}c@{}} Target network update frequency \\ \end{tabular} & Every $50$ steps \\
\hline
\end{tabular}
\vspace{-15pt}
\label{tab_sim_par}
\end{center}
\end{table}

\vspace{-5pt}

\subsection{Scenario 1: Single-Channel Setting}
\begin{figure}[t!]\centering
\centerline{\includegraphics[width=3.5in]{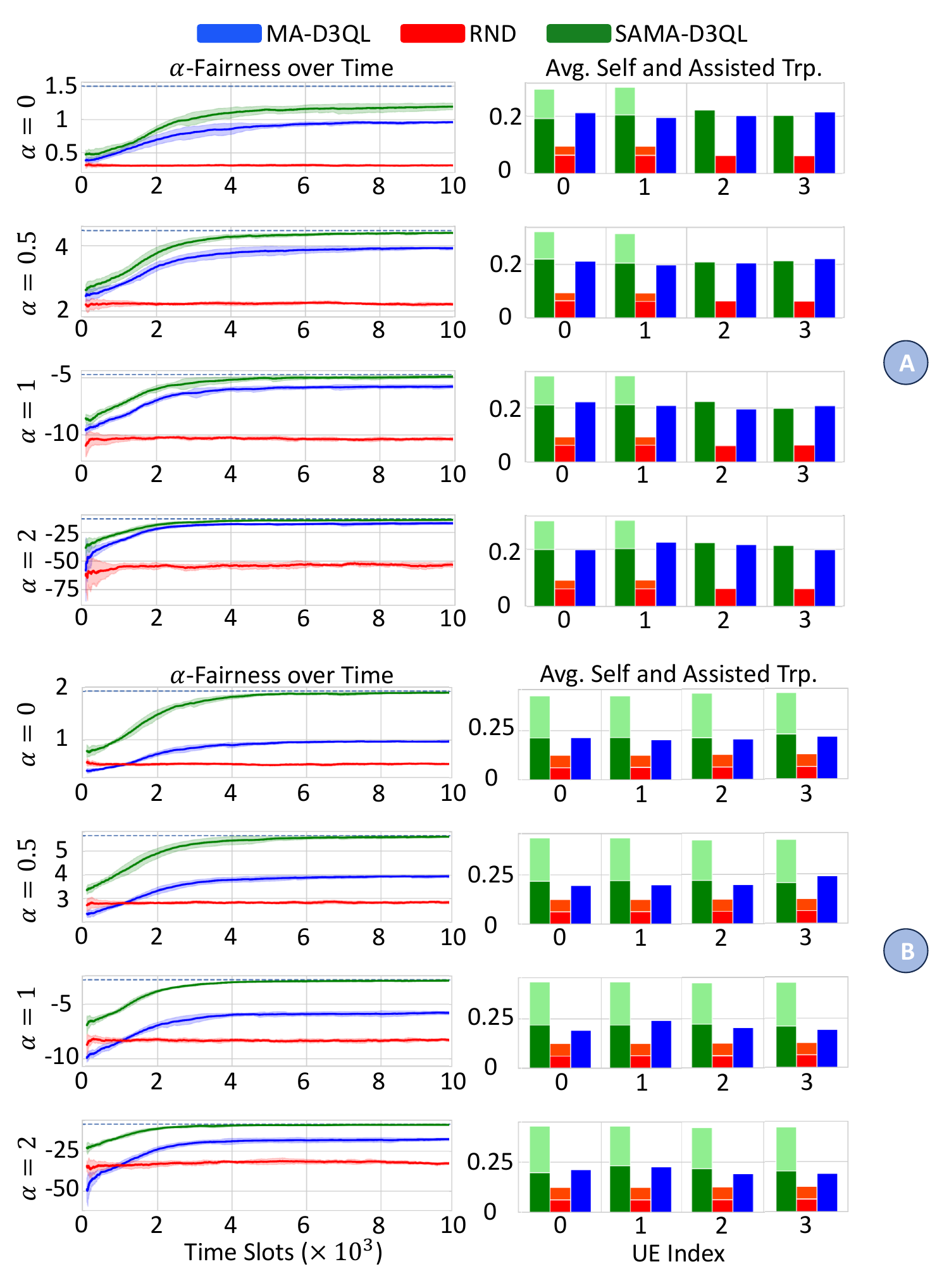}}
\vspace{-10pt}
\caption{Objective functions over time and all-time average UE throughputs in the first scenario, encompassing experiments A and B, comparing the SAMA-D3QL, MA-D3QL, and RND (random agents) algorithms. The reported results are based on averaging the outcomes of 10 rounds of simulations. It's noteworthy that, when present, the lighter segment of each UE throughput signifies the assisted throughput for that particular UE.}
\label{fig_snr01}
\vspace{-15pt}
\end{figure}

In this scenario, we conduct a comparative analysis of the algorithms using two predefined UE-segment association matrices. The first matrix (Fig. \ref{fig_snr01}-A) is the one depicted in Fig. \ref{fig_sample_net} with optimal points stated in Table. \ref{sample_net_optimal}. The second matrix (Fig. \ref{fig_snr01}-B) associates two segments with each pair of four users, represented as $A = \big[[1, 1, 0, 0],[1, 1, 0, 0],[0, 0, 1, 1],[0, 0, 1, 1]\big]$. Fig. \ref{fig_snr01} illustrates the evolution of objective functions over time on the left side and the average UE throughputs over all-time on the right side. The lighter portion of each UE throughput, if present, indicates the assisted throughput of that UE. Across different values of $\alpha$, SAMA-D3QL consistently and significantly outperforms MA-D3QL in both network configurations converging to the optimal solution (as denoted by the dotted line). It's worth noting that in the single-channel setup, optimal values are obtained through exhaustive searches over self-throughputs. However, this calculation becomes infeasible in the multi-channel network, where UEs sharing segments may transmit simultaneously.

\vspace{-5pt}

\subsection{Scenario 2: Multi-channel Setting}
\begin{figure*}[!t]
\centerline{\includegraphics[width=6in]{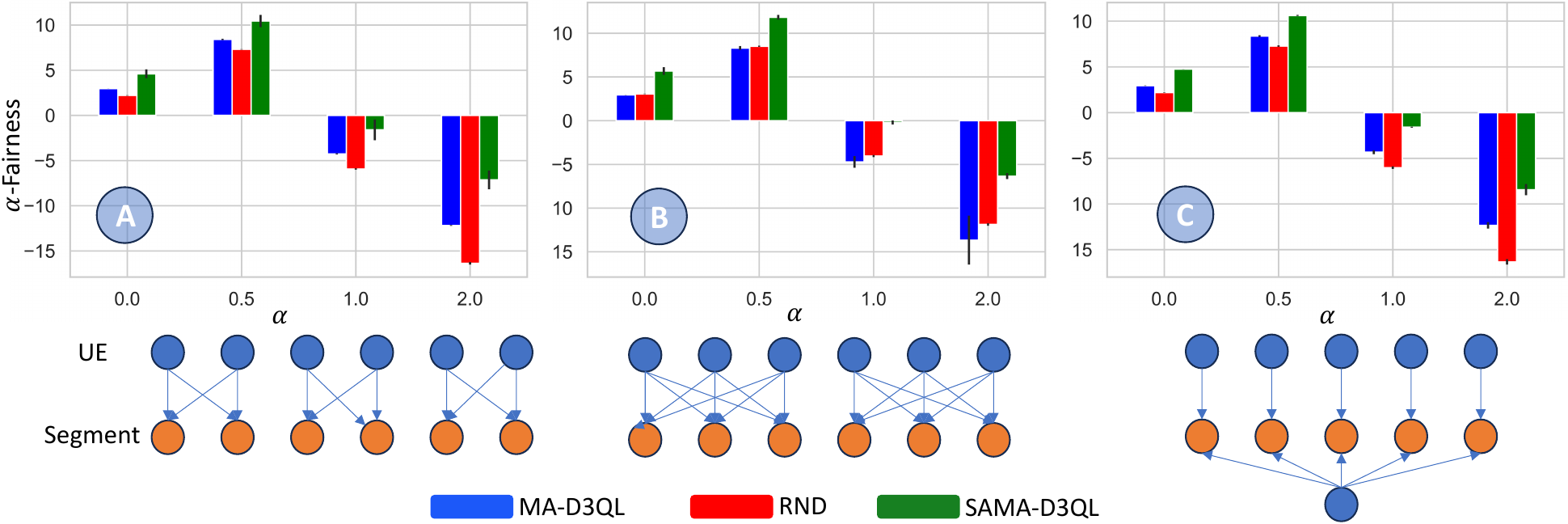}}
\vspace{-5pt}
\caption{The average value of objective functions for the last 1000 time slots (of 10k time slots) in experiments A, B, and C. The UE-segment association matrix for each experiment is illustrated. Results are the average of 10 rounds of simulations.}
\label{fig_snr02}
\vspace{-15pt}
\end{figure*}

In this scenario, we have conducted three experiments with the objective of assessing the impact of fairness levels (quantified by the parameter $\alpha$) and the sparsity of association matrices on algorithm effectiveness. The experimental setup entails six UEs contending for access to three channels, with varying trade-offs between utilization and fairness. In addition, three different association matrices, each characterized by a distinct level of sparsity, are utilized. The results depicted in Fig. \ref{fig_snr02} highlight the superior performance of SAMA-D3QL agents across all experiments. However, the extent of this performance advantage varies depending on the specific association matrix and the chosen fairness level. Intriguingly, random agents achieve comparable results to MA-D3QL agents, underscoring the potential of semantic awareness in enhancing the efficiency and efficacy of communication systems. Notably, in this particular scenario with multiple channels, random agents exhibit better performance compared to the previous scenario, likely due to the increased channel diversity. Lastly, the enhanced performance of SAMA-D3QL is particularly notable in scenarios with denser association matrices, such as in topology B. This implies that semantic-aware multiple access could provide increased advantages in situations with more correlated traffic.

\section{Conclusion}\label{s_con}
This paper introduced a novel semantic-aware problem formulation and algorithmic approach for addressing the challenges of multiple access in distributed and dynamic 6G-based applications. The core objective of this research was to strike a balance between maximizing network utilization and ensuring fairness, all while considering the inherent data correlations among users. To achieve this goal, we devised a structured and adaptable formal framework that leverages the semantic relationships between users. Additionally, we introduced the SAMA-D3QL algorithm, which demonstrates a remarkable capacity to significantly enhance the $\alpha$-fairness of the system compared to semantic-oblivious approaches. Importantly, this approach relies on decentralized multi-agent Q-learning, enabling UEs to make autonomous decisions regarding wireless spectrum access without requiring global knowledge of the system. This advancement holds the potential to bring future federated, dynamically evolving applications one step closer to real-world deployment.

Our prospective future endeavors will involve expanding this framework to encompass more realistic and practical scenarios, including those with time-varying and non-binary UE-segment association matrices, as well as variable packet-size scenarios. Implementing more advanced MADRL algorithms with reduced communication overhead would also enhance the practicality of our proposed scheme. In addition, it would be advantageous to assess the framework's performance in real-world settings where approximate semantic correlations are prevalent.

\begin{appendices}
\section{General Setting of the Problem} \label{app_a}
Throughout this paper, we have operated under the assumption that the reception of multiple duplicates of a segment by the SBS does not enhance throughput. As a result, we have employed the minimum function in \eqref{segment_indicator} and subsequently dissected user throughput into self and assisted components. In a more generalized context, where the association matrix $\mathbb{A}$ is non-binary or the channel exhibits imperfections, we must reformulate \eqref{segment_indicator} in the following manner:
\begin{equation}\label{segment_indicator_general}
y_{k}^{t} \triangleq \textbf{g} \bigg( \{ a_{j,k}^{t} \cdot z_{j}^{t} \ | \ j \in \mathbb{N} \} \bigg),
\end{equation}
where $\textbf{g}(.)$ represents a function that determines the successful reception of a segment by the SBS based on all received signals from the UEs. Consequently, we should employ \eqref{ue_throughput} (instead of \eqref{ue_throughput_decomposed}) to formulate C4 in \eqref{problem}.
\end{appendices}

\section*{Acknowledgment}
This research work was also conducted at ICTFICIAL Oy. It is partially supported by the European Union’s Horizon 2020 Research and Innovation Program through the aerOS project under Grant No. 101069732; the Business Finland 6Bridge 6Core project under Grant No. 8410/31/2022; the European Union’s HE research and innovation programe HORIZON-JUSNS-2022 under the 6GSandbox project (Grant No. 101096328); and and the Research Council of Finland 6G Flagship Programme under Grant No. 346208. The paper reflects only the authors’ views. The Commission is not responsible for any use that may be made of the information it contains.

\bibliographystyle{IEEEtran}
\bibliography{IEEEabrv,Bibliography}

\end{document}